\documentclass[aps,twocolumn]{revtex4}
\usepackage{url}
\usepackage[colorlinks]{hyperref}
\hypersetup{%
    plainpages=true,
    breaklinks=true,
    hypertexnames=false,
    pageanchor=true,
    colorlinks=true,
    linkcolor={blue},
    citecolor={blue},
    urlcolor={blue},
    anchorcolor={black}
}

\usepackage{amsmath}
\usepackage{graphicx}
\usepackage{subfigure}

\def\yyy#1{{{\color{black} #1 \color{black}}}}

\begin{document}

\title{Sender-controlled measurement-device-independent multiparty quantum communication}
\author{Yuyan Wei$^{1}$,  Siying Wang$^{1}$, Yajing Zhu$^{1}$, and  Tao Li$^{1,2}$\footnote{tao.li@njust.edu.cn}
}
\affiliation{$^{1}$ School of Science, Nanjing University of Science and Technology, Nanjing 210094, China\\
$^{2}$ MIIT Key Laboratory of Semiconductor Microstructure, Nanjing University of Science and Technology, Nanjing 210094, China
}
\date{\today }
\newpage

\begin{abstract}
Multiparty quantum communication is an important branch of quantum networks. It enables private information transmission with information-theoretic security among legitimate parties.  We propose a sender-controlled measurement-device-independent multiparty quantum communication protocol. The sender Alice  divides a private message into several parts and delivers them to different receivers for secret sharing with imperfect measurement devices and  untrusted ancillary nodes. Furthermore,  Alice acts as an active controller and checks the security of quantum channels and the reliability of each receiver before she encodes her private message for secret sharing, which makes the protocol convenient for multiparity quantum communication.\\
\end{abstract}

\maketitle

\section{Introduction} \label{sec1}
Quantum communication, such as quantum key distribution (QKD)~\cite{Gisin2002RMP,cui2019measurement,Shang2020oneway,Yan2020Measurement,Zhang2020QKD}, quantum secure direct communication (QSDC)~\cite{long2002theoretically,deng2003two,wang2005quantum,hu2016experimental,zhang2017quantum,chen2018three,Li2020high-capacity,Li2020QSDC,Ye2021Generic}, and quantum secret sharing (QSS)~\cite{Hillery1999,Cleve1999How,Tittel2001,Chen2005Experimental,gao2005deterministic}, provides an absolutely secure technique that transmits private information between legitimate parties. Bennett and Brassard proposed the first quantum communication protocol~\cite{bb84}, enabling two parties to share private key for encryption. In principle, eavesdroppers attacking on a quantum communication process introduce perturbations that inevitably reveal their interception~\cite{Gisin2002RMP}. However, practical apparatuses with imperfect functions may have loopholes for  side-channel attacks on quantum cryptography~\cite{lo2014secure,Xu2020Secure,Li2020MDIreview}. In principle, the device independent architecture based on the violation of Bell inequality can remove all side-channel attacks on  quantum cryptography using non-ideal devices~\cite{Acin2007Device,Lim2013Device,ZHOU202012}, whereas it requires a loophole-free Bell test~\cite{Xu2020Secure}. Measurement-device-independent~(MDI) architecture~\cite{lo2012measurement} provides a simplified strategy for removing serious side-channel attacks on practical measurement apparatuses by using postselected entanglement. For instance, two legitimate parties with practical measurement apparatuses  can share private key by MDI-QKD~\cite{lo2014secure,Xu2020Secure} and directly exchange classical messages over quantum channels by MDI-QSDC~\cite{niu2018measurement,zhou2020measurement,gao2019long,Zou2020Measurement}.

Multiparty quantum communication involves more than two nodes of a network and directly transmits private
information among them without any classical relay~\cite{wehner2018quantum,Qin2016Controllable,Qi202115user}. QSS is a typical multiparty quantum communication  protocol~\cite{Hillery1999,Cleve1999How,Tittel2001,Chen2005Experimental,gao2005deterministic}. It splits a private message into several parts and sends each part to one party; the message can only be reconstructed through cooperation between all parties. In 1999, Hillery, Bu\v{z}ck and Berthiaume~\cite{Hillery1999} proposed the first QSS protocol  using a maximally entangled three-particle  Greenberger-Horne-Zeilinger~(GHZ) state; Karlsson et al.~\cite{Karlsson1999} proposed a QSS protocol using two-particle quantum entanglement. In 2004, Xiao et al.~\cite{Xiao2004QSS}  proposed a high-efficient QSS protocol, increasing its efficiency to asymptotically $100\%$  by  properly choosing measurement bases~\cite{lo2005efficient,xue2017efficient}. Subsequently, some interesting QSS protocols were proposed~\cite{LI2004420,Zhang2005QSS, ZHANG200560,
yang2018detector,Huang2019Securing,
Xiang2017Multipartite,Kogias2017Unconditional,Habibidavijani_2019}.

In 2015, Fu at al.\cite{Fu2015} proposed  the first MDI-QSS protocol with  postselected GHZ states and closed  loopholes introduced by practical measurement apparatuses. Recently, Gao et al.~\cite{gao2020deterministic} presented a deterministic MDI-QSS~(DMDI-QSS) protocol and  removed the requirement of basis reconciliation in the  MDI architecture~\cite{lo2014secure,Xu2020Secure,Li2020MDIreview}.
In this protocol, the sender Alice produces entangled photon pairs and sends one photon from each pair to David, whereas the receivers (Bob and Charlie) randomly produce single photons in an eigenstate of  $\sigma_X$ or  $\sigma_Y$ basis, and also send their photons to David for GHZ-state analysis. The GHZ-state analyzer in David's node~\cite{Pan1998GHZanalyzer,Lu2009swap,Kop2007linear}, in principle, performs a teleportation operation and correlates the states of photons kept in Alice's node to that prepared by the receivers.
The receivers check the security of DMDI-QSS before information encoding; and some of receivers can cheat one receiver by a particular strategy~\cite{Deng2005ImproveQSS,Yang2021Participant,Yang2021stronger}.
Therefore, it requires at least two faithful receivers to guarantee the security of DMDI-QSS against the participant attack~\cite{gao2020deterministic}.

Here we present a sender-controlled MDI-QSS protocol, in which the sender performs as an active controller~\cite{ting2005controlled} and can detect the attack from either outside eavesdropper or unfaithful receiver~(i.e., participant attack) by cooperating with receivers. Single photons with random polarization inserted into entangled photon pairs are used for security check~\cite{niu2018measurement,zhou2020measurement,gao2019long}. Once a postselected GHZ state is produced  by  ancillary David from individual single photons, exclusive correlations are created among photon states prearranged by the sender and receivers. After receiving  the photon states published by all  receivers,  the sender can perform security check based on the postselected multiphoton entanglement rather than the multiphoton quantum teleportation that requires ancillary measurements~\cite{gao2020deterministic}. Therefore, the sender acts as an active controller and encodes private information after ascertaining the security of quantum channels and the reliability of each receiver,  which enables the receivers to  share private messages over  quantum channels and makes this protocol  convenient for practical secret sharing.

\section{Three-party sender-controlled MDI-QSS protocol}

\begin{figure}
	\centering
	\includegraphics[scale=0.4]{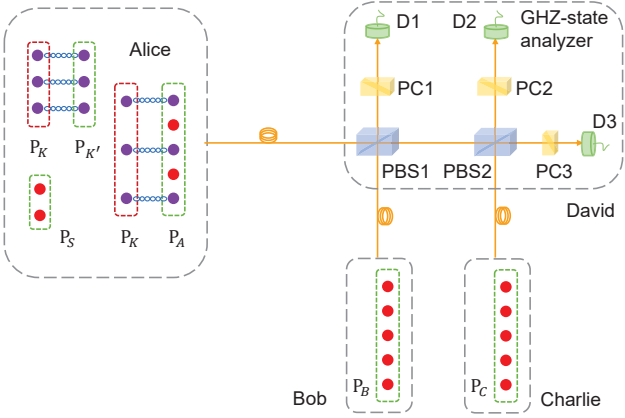}
	\caption{Schematic diagram of sender-controlled MDI-QSS protocol. PBS1 and PBS2 are two polarizing beam splitters that transmit photons with horizontal polarization $|0\rangle$ and reflect {photons} with vertical polarization $|1\rangle$; PC is a polarization controller and completes a $\pi/4$ rotation of the polarization; \yyy{D$1$, D$2$, and D$3$} are single-photon detectors, \yyy{which can distinguish between two polarizations $\vert0\rangle$ and $\vert1\rangle$.}
	} \label{fig1}
\end{figure}

A sender-controlled MDI-QSS protocol among three legitimate parties is shown in Fig.~\ref{fig1}. The sender Alice prepares a sequence of  entangled photon pairs and divides them into two sequences. She sends one photon sequence to ancillary David after she randomly inserts some single photons with random  polarization in it. Bob and Charlie prepare
single photons with random polarization and send them to David as well. David performs GHZ-state analysis of each three photons with the same order and heralds exclusive correlations among three legitimate parties by either postselected entanglement generation or quantum teleportation. These two cases will be used for security check and secret splitting, respectively. Specifically, the three-party sender-controlled MDI-QSS proposal is carried out in the following steps.

Step (1) Photon-sequence preparation. Alice prepares $k_1$ entangled photon pairs in  the Bell state  $\vert \varphi^-\rangle={(\vert 01\rangle -\vert 10\rangle)}/{\sqrt{2}}$, where  $|0\rangle$~($|1\rangle$)  represents a  horizontally~(vertically) polarized photon,
and divides them into two sequences $P_K$ and $P_{K^{\prime}}$ by choosing one photon from each pair and rearranging them with the same order as in photon pair sequence~\cite{long2002theoretically,deng2003two,wang2005quantum}.
She also prepares a sequence $P_S$ consisting of $k_2$ single photons which are randomly prepared in either $\sigma_X$~[i.e., $\vert\pm x\rangle=(\vert0\rangle\pm\vert1\rangle)/\sqrt{2}$] or $\sigma_Y$~[i.e., $|\pm y\rangle= (\vert0\rangle\pm  i\vert1\rangle)/\sqrt{2}$] basis, and randomly inserts them into $P_{K^{\prime}}$ to reform a new sequence $P_A$ consisting of  $n=k_1+k_2$ photons.
Meanwhile, Bob and Charlie prepare $n$-photon {sequences} $P_B$ and $P_C$ with each photon randomly initialized in either $\sigma_X$ or  $\sigma_Y$ basis, respectively.

Step (2) Photon-sequence transmission. Alice, Bob, and Charlie send their photon sequences $P_A$, $P_B$, and $P_C$ to the untrusted ancillary David to perform a GHZ-state analysis of each three photons with the same order.  The  GHZ-state analyzer can project the incoming photons into one of three-photon GHZ states that can be described as
\begin{equation}
	\begin{split}
		\vert\Phi\rangle_{000}=\frac{1}{\sqrt{2}}({\vert000\rangle+\vert111\rangle}), \quad \vert\Phi\rangle_{001}=\frac{1}{\sqrt{2}}({\vert000\rangle-\vert111\rangle}),\\
		\vert\Phi\rangle_{010}=\frac{1}{\sqrt{2}}({\vert001\rangle+\vert110\rangle}), \quad \vert\Phi\rangle_{011}=\frac{1}{\sqrt{2}}({\vert001\rangle-\vert110\rangle}),\\
		\vert\Phi\rangle_{100}=\frac{1}{\sqrt{2}}({\vert010\rangle+\vert101\rangle}), \quad \vert\Phi\rangle_{101}=\frac{1}{\sqrt{2}}({\vert010\rangle-\vert101\rangle}),\\
		\vert\Phi\rangle_{110}=\frac{1}{\sqrt{2}}({\vert011\rangle+\vert100\rangle}), \quad \vert\Phi\rangle_{111}=\frac{1}{\sqrt{2}}({\vert011\rangle-\vert100\rangle}).\\
	\end{split}\label{eq1}
\end{equation}
In practice, a GHZ-state analyzer can only identify two GHZ states $\vert\Phi\rangle_{000}$ and $\vert\Phi\rangle_{001}$ when using  linear optical elements and single-photon detectors~\cite{Pan1998GHZanalyzer,Lu2009swap,Kop2007linear}, shown in Fig.~\ref{fig1}. \yyy{PBS1 and PBS2 are two polarizing beam splitters that transmit photons with horizontal polarization $|0\rangle$ and reflect {photons} with vertical polarization $|1\rangle$; PC is a polarization controller and completes a $\pi/4$ rotation of the polarization; D$1$, D$2$, and D$3$ are single-photon detectors and can distinguish between two polarizations $\vert0\rangle$ and $\vert1\rangle$, reporting 0 and 1, respectively. Therefore, the state $\vert\Phi\rangle_{000}$ leads to a coincidence event of  three-detector clicks $000$, $011$, $101$, or $110$, while the state $\vert\Phi\rangle_{001}$ results in an event $001$, $010$, $100$, or $111$~\cite{Lu2009swap}. Although the success probability of this analysis is $1/4$,} it is enough for security check and secret sharing of our three-party sender-controlled MDI-QSS protocol.

When a photon $k'$ of sequence $P_A$ entering the GHZ-state analyzer originates from $P_{K'}$ and entangles with the photon $k$ of  sequence $P_K$, the GHZ-state analysis can correlate the  {state} of photon $k$ kept in Alice's node and the single-photon states $\vert i\rangle$ and $\vert  j\rangle$  prepared by Bob and Charlie with $i, j\in\{+x,-x,+y,-y\}$. In general, the states of photons $k$ kept in Alice's node will be projected into an eigenstate of $\sigma_X$~($\sigma_Y$) if Bob and Charlie prepare their photons in basis \yyy{$\sigma_X\otimes\sigma_Y$} or $\sigma_Y\otimes\sigma_X$~($\sigma_X\otimes\sigma_X$ or $\sigma_Y\otimes\sigma_Y$)~\cite{gao2020deterministic} \yyy{and David reports successful GHZ-state measurements on photons sent to him}, shown in Tables~\ref{Table1} and \ref{Table2}. \yyy{$\alpha$ is the number of $\sigma_Y$ basis that Bob and Charlie use for single-photon preparation. $\beta$ is the number of states $\vert-x\rangle$ and $\vert-y\rangle$ that Bob and Charlie prepared. $\gamma_j\leq \text{Min}\{\alpha,1\}$ is the number of state $|1\rangle$ of photons that are prepared in the $\sigma_Y$ basis in each $\vert\omega\rangle_j$. Here $\vert\omega\rangle_j$ with $j=1,2,3,4$ are four  state vectors of the $\sigma_Z\otimes\sigma_Z$ basis, the superposition of which consists the two-photon state $\vert i\rangle\otimes\vert  j\rangle$. For instance, David announces a GHZ state $\vert\Phi\rangle_{000}$, Bob and Charlie prepare their photons in the state $\vert+x\rangle\otimes\vert+x\rangle$, this corresponds to the Case 1 in Table~\ref{Table1}~($\alpha=0$, $\beta=0$) with $\gamma_j=0$ and $\alpha-2\gamma_j=0$, and the  corresponding photon in sequence $P_K$ is projected into the state $\vert-x\rangle$; David announces a GHZ state $\vert\Phi\rangle_{001}$, Bob and Charlie prepare their photons in the state $\vert+x\rangle\otimes\vert-y\rangle$,  this corresponds to the Case 4 in Table~\ref{Table2}~($\alpha=1$, $\beta=1$) with $\gamma_j=0$~($\gamma_j=1$) and $\alpha-2\gamma_j=1$~($\alpha-2\gamma_j=-1$), and the corresponding photon in sequence $P_K$ is projected into the state $\vert+y\rangle$~($\vert-y\rangle$).} Alice can encode her information by applying local operations on photons kept in her node, and then sends them to receivers for  secret sharing.

When a photon $s$ of sequence $P_A$ entering the GHZ-state analyzer originates from $P_S$, it is separable from all other photons. The product state of three photons entering GHZ-state analyzer can be described as a superposition of three-photon GHZ states. The GHZ-state analyzer can only report one GHZ state of the superposition each time, referred to as postselected entanglement. In general, when Alice, Bob, and Charlie prepare their photons in the bases with even $\sigma_Y$, such as $\sigma_X\otimes\sigma_X\otimes\sigma_X$, $\sigma_X\otimes\sigma_Y\otimes\sigma_Y$,
$\sigma_Y\otimes\sigma_X\otimes\sigma_Y$,
$\sigma_Y\otimes\sigma_Y\otimes\sigma_X$, any three-photon product state will be a superposition of four GHZ states, whereas it never simultaneously contains \yyy{states $\vert\Phi\rangle_{b_1b_20}$ and $\vert\Phi\rangle_{b_1b_21}$ with relative phase differences $0$ and $\pi$ between two components consisting them, respectively. Here the subscripts $b_1, b_2\in\{0,1\}$ of $\vert\Phi\rangle_{b_1b_20}$ are identical to that of $\vert\Phi\rangle_{b_1b_21}$.} For instance, when Alice, Bob, and Charlie prepare their single photons in  $\sigma_X\otimes\sigma_X\otimes\sigma_X$ basis, eight three-photon product states can be described in the GHZ-state basis as follows:
\begin{equation}
	\begin{split}
		&\vert{+x+\!x+\!x}\rangle
		\!=\!\frac{1}{2}(\vert\Phi\rangle_{000}\!+\!\vert\Phi\rangle_{010}\!+\!\vert\Phi\rangle_{100}\!+\!\vert\Phi\rangle_{110}),
		\\
		&\vert{+x+\!x-\!x}\rangle
		\!=\!\frac{1}{2}(\vert\Phi\rangle_{001}\!-\!\vert\Phi\rangle_{011}\!+\!\vert\Phi\rangle_{101}\!-\!\vert\Phi\rangle_{111}),
		\\
		&\vert{+x-\!x+\!x}\rangle
		\!=\!\frac{1}{2}(\vert\Phi\rangle_{001}\!+\!\vert\Phi\rangle_{011}\!-\!\vert\Phi\rangle_{101}\!-\!\vert\Phi\rangle_{111}),
		\\
				&\vert{-x+\!x+\!x}\rangle
		\!=\!\frac{1}{2}(\vert\Phi\rangle_{001}\!+\!\vert\Phi\rangle_{011}\!+\!\vert\Phi\rangle_{101}\!+\!\vert\Phi\rangle_{111}),
		\\
&\vert{+x-\!x-\!x}\rangle
		\!=\!\frac{1}{2}(\vert\Phi\rangle_{000}\!-\!\vert\Phi\rangle_{010}\!-\!\vert\Phi\rangle_{100}\!+\!\vert\Phi\rangle_{110}),
		\\
		&\vert{-x+\!x-\!x}\rangle
		\!=\!\frac{1}{2}(\vert\Phi\rangle_{000}\!-\!\vert\Phi\rangle_{010}\!+\!\vert\Phi\rangle_{100}\!-\!\vert\Phi\rangle_{110}),
		\\
		&\vert{-x-\!x+\!x}\rangle
		\!=\!\frac{1}{2}(\vert\Phi\rangle_{000}\!+\!\vert\Phi\rangle_{010}\!-\!\vert\Phi\rangle_{100}\!-\!\vert\Phi\rangle_{110}),
		\\
		&\vert{-x-\!x-\!x}\rangle
		\!=\!\frac{1}{2}(\vert\Phi\rangle_{001}\!-\!\vert\Phi\rangle_{011}\!-\!\vert\Phi\rangle_{101}\!+\!\vert\Phi\rangle_{111}).
	\end{split}\;\;\;\;\;\label{eq.2}
\end{equation}

When Alice, Bob, and Charlie prepare their photons in the basis $\sigma_X\otimes\sigma_Y\otimes\sigma_Y$, any three-photon product state will be referred to as a superposition of four GHZ states similar to the aforementioned case as follows:
\begin{equation}
	\begin{split}
		&\vert+\!x+\!y+\!y\rangle
		\!=\!\frac{1}{2}(\vert\Phi\rangle_{001}\!+\!i\vert\Phi\rangle_{010}\!+\!i\vert\Phi\rangle_{100}\!-\!\vert\Phi\rangle_{111}),
		\\
		&\vert+\!x+\!y-\!y\rangle
		\!=\!\frac{1}{2}(\vert\Phi\rangle_{000}\!-\!i\vert\Phi\rangle_{011}\!+\!i\vert\Phi\rangle_{101}\!+\!\vert\Phi\rangle_{110}),
		\\
		&\vert+\!x-\!y+\!y\rangle
		\!=\!\frac{1}{2}(\vert\Phi\rangle_{000}\!+\!i\vert\Phi\rangle_{011}\!-\!i\vert\Phi\rangle_{101}\!+\!\vert\Phi\rangle_{110}),
		\\
		&\vert-\!x+\!y+\!y\rangle
		\!=\!\frac{1}{2}(\vert\Phi\rangle_{000}\!+\!i\vert\Phi\rangle_{011}\!+\!i\vert\Phi\rangle_{101}\!-\!\vert\Phi\rangle_{110}),
		\\
		&\vert+\!x-\!y-\!y\rangle
		\!=\!\frac{1}{2}(\vert\Phi\rangle_{001}\!-\!i\vert\Phi\rangle_{010}\!-\!i\vert\Phi\rangle_{100}\!-\!\vert\Phi\rangle_{111}),
		\\
		&\vert-\!x+\!y-\!y\rangle
		\!=\!\frac{1}{2}(\vert\Phi\rangle_{001}\!-\!i\vert\Phi\rangle_{010}\!+\!i\vert\Phi\rangle_{100}\!+\!\vert\Phi\rangle_{111}),
		\\
		&\vert-\!x-\!y+\!y\rangle
		\!=\!\frac{1}{2}(\vert\Phi\rangle_{001}\!+\!i\vert\Phi\rangle_{010}\!-\!i\vert\Phi\rangle_{100}\!+\!\vert\Phi\rangle_{111}),
		\\
		&\vert-\!x-\!y-\!y\rangle
		\!=\!\frac{1}{2}(\vert\Phi\rangle_{000}\!-\!i\vert\Phi\rangle_{011}\!-\!i\vert\Phi\rangle_{101}\!-\!\vert\Phi\rangle_{110}),
	\end{split}\label{eq.3}
\end{equation}

Similarly, when Alice, Bob, and Charlie prepare their photons in the basis $\sigma_Y\otimes\sigma_X\otimes\sigma_Y$ or $\sigma_Y\otimes\sigma_Y\otimes\sigma_X$, any three-photon product state will also be referred to as a superposition of four GHZ states. For these two cases, any three-photon product state can be, respectively, described as
\begin{equation}
	\begin{split}
		&\vert+\!y+\!x+\!y\rangle
		\!=\!\frac{1}{2}(\vert\Phi\rangle_{001}\!+\!i\vert\Phi\rangle_{010}\!+\!\vert\Phi\rangle_{101}\!+\!i\vert\Phi\rangle_{110}),
		\\
		&\vert+\!y+\!x-\!y\rangle
		\!=\!\frac{1}{2}(\vert\Phi\rangle_{000}\!-\!i\vert\Phi\rangle_{011}\!+\!\vert\Phi\rangle_{100}\!-\!i\vert\Phi\rangle_{111}),
		\\
		&\vert+\!y-\!x+\!y\rangle
		\!=\!\frac{1}{2}(\vert\Phi\rangle_{000}\!+\!i\vert\Phi\rangle_{011}\!-\!\vert\Phi\rangle_{100}\!-\!i\vert\Phi\rangle_{111}),
		\\
		&\vert-\!y+\!x+\!y\rangle
		\!=\!\frac{1}{2}(\vert\Phi\rangle_{000}\!+\!i\vert\Phi\rangle_{011}\!+\!\vert\Phi\rangle_{100}\!+\!i\vert\Phi\rangle_{111}),
		\\
		&\vert+\!y-\!x-\!y\rangle
		\!=\!\frac{1}{2}(\vert\Phi\rangle_{001}\!-\!i\vert\Phi\rangle_{010}\!-\!\vert\Phi\rangle_{101}\!+\!i\vert\Phi\rangle_{110}),
		\\
		&\vert-\!y+\!x-\!y\rangle
		\!=\!\frac{1}{2}(\vert\Phi\rangle_{001}\!-\!i\vert\Phi\rangle_{010}\!+\!\vert\Phi\rangle_{101}\!-\!i\vert\Phi\rangle_{110}),
		\\
		&\vert-\!y-\!x+\!y\rangle
		\!=\!\frac{1}{2}(\vert\Phi\rangle_{001}\!+\!i\vert\Phi\rangle_{010}\!-\!\vert\Phi\rangle_{101}\!-\!i\vert\Phi\rangle_{110}),
		\\
		&\vert-\!y-\!x-\!y\rangle
		\!=\!\frac{1}{2}(\vert\Phi\rangle_{000}\!-\!i\vert\Phi\rangle_{011}\!-\!\vert\Phi\rangle_{100}\!+\!i\vert\Phi\rangle_{111}),
	\end{split}\label{eq.4}
\end{equation}
or
\begin{equation}
	\begin{split}
		&\vert+\!y+\!y+\!x\rangle
		\!=\!\frac{1}{2}(\vert\Phi\rangle_{001}\!+\!\vert\Phi\rangle_{011}\!+\!i\vert\Phi\rangle_{100}\!+\!i\vert\Phi\rangle_{110}),
		\\
		&\vert+\!y+\!y-\!x\rangle
		\!=\!\frac{1}{2}(\vert\Phi\rangle_{000}\!-\!\vert\Phi\rangle_{010}\!+\!i\vert\Phi\rangle_{101}\!-\!i\vert\Phi\rangle_{111}),
		\\
		&\vert+\!y-\!y+\!x\rangle
		\!=\!\frac{1}{2}(\vert\Phi\rangle_{000}\!+\!\vert\Phi\rangle_{010}\!-\!i\vert\Phi\rangle_{101}\!-\!i\vert\Phi\rangle_{111}),
		\\
		&\vert-\!y+\!y+\!x\rangle
		\!=\!\frac{1}{2}(\vert\Phi\rangle_{000}\!+\!\vert\Phi\rangle_{010}\!+\!i\vert\Phi\rangle_{101}\!+\!i\vert\Phi\rangle_{111}),
		\\
		&\vert+\!y-\!y-\!x\rangle
		\!=\!\frac{1}{2}(\vert\Phi\rangle_{001}\!-\!\vert\Phi\rangle_{011}\!-\!i\vert\Phi\rangle_{100}\!+\!i\vert\Phi\rangle_{110}),
		\\
		&\vert-\!y+\!y-\!x\rangle
		\!=\!\frac{1}{2}(\vert\Phi\rangle_{001}\!-\!\vert\Phi\rangle_{011}\!+\!i\vert\Phi\rangle_{100}\!-\!i\vert\Phi\rangle_{110}),
		\\
		&\vert-\!y-\!y+\!x\rangle
		\!=\!\frac{1}{2}(\vert\Phi\rangle_{001}\!+\!\vert\Phi\rangle_{011}\!-\!i\vert\Phi\rangle_{100}\!-\!i\vert\Phi\rangle_{110}),
		\\
		&\vert-\!y-\!y-\!x\rangle
		\!=\!\frac{1}{2}(\vert\Phi\rangle_{000}\!-\!\vert\Phi\rangle_{010}\!-\!i\vert\Phi\rangle_{101}\!+\!i\vert\Phi\rangle_{111}).
	\end{split}\label{eq.5}
\end{equation}

However, when  Alice, Bob, and Charlie prepare their photons in the bases with odd $\sigma_Y$, such as $\sigma_Y\otimes\sigma_X\otimes\sigma_X$, $\sigma_X\otimes\sigma_Y\otimes\sigma_X$,
$\sigma_X\otimes\sigma_X\otimes\sigma_Y$,
$\sigma_Y\otimes\sigma_Y\otimes\sigma_Y$, any three-photon product state will be a superposition of eight GHZ states with an equal probability of $1/8$.
For instance, the  three-photon product state $\vert+\!y+\!x+\!x\rangle$ prepared in $\sigma_Y\otimes\sigma_X\otimes\sigma_X$ basis can be described in the three-photon GHZ-state basis as follows:
\begin{equation}
	\begin{split}
		\vert+\!y+\!x+\!x\rangle
		\!=\!&\frac{1}{4}[(1\!+\!i)(\vert\Phi\rangle_{000}\!+\!\vert\Phi\rangle_{010}\!+\!\vert\Phi\rangle_{100}\!+\!\vert\Phi\rangle_{110})\\
		&+(1\!-\!i)(\vert\Phi\rangle_{001}\!+\!\vert\Phi\rangle_{011}\!+\!\vert\Phi\rangle_{101}\!+\!\vert\Phi\rangle_{111})].
	\end{split}
\end{equation}
Therefore, for any three-photon product state, it can be described in  the GHZ-state basis by a deterministic superposition. Once the three photons are prepared in the  {bases} with even $\sigma_Y$, the GHZ-state analysis of them will report one of four GHZ states, whereas  the GHZ-state analyzer will report one of eight GHZ states when  they are  prepared in the bases with odd $\sigma_Y$. The former case will be used for security check, because any eavesdropper will distort the GHZ-state analysis and can lead to a GHZ state that is not contained in the superposition, shown in
Eqs.~(\ref{eq.2})-(\ref{eq.5}).

Step (3) Security check. David announces the results of GHZ-state analysis through public channels, together with the corresponding position of the photons. Alice, Bob, and Charlie keep the raw data of successful GHZ-state analysis and discard the rest. To detect eavesdropping, Alice publishes the  positions of single photons originating from $P_s$ in sequence $P_A$ and asks Bob and Charlie to publish their photon states with the same order in $P_B$ and $P_C$. By using the cases in which they have prepared three photons in bases with even $\sigma_Y$, Alice will calculate the error rate of the photon-sequence transmission and judges whether there is an eavesdropper or unfaithful participant intercepting the quantum channel. When the error rate is larger than the threshold, she will stop the communication process and asks Bob and Charlie to restart from step (1); otherwise, she
moves to the next procedure for message encoding.

Step (4) Message encoding. After the security check is passed, Alice encodes her secret message on the photons in a modified sequence $P_{K^{\prime\prime}}$ which is a subsequence of $P_K$ and their counterparts have led to a faithful GHZ-state analysis. Specifically, if the bit value is $0$, she does not perform any operation on the corresponding photon, if the bit value is $1$, she performs a unitary operation $U=\vert0\rangle\langle0\vert-\vert1\rangle\langle1\vert$ on the corresponding photon, which completes the transformation $\vert +x \rangle\leftrightarrow\vert-x \rangle$ and $\vert +y \rangle\leftrightarrow\vert-y \rangle$. Meanwhile, Alice encodes some sampling bits together with the message encoding for integrity check.

Step (5) Message decoding. In practice, there are two methods for Bob and Charlie to decode the message sent by Alice: (I) Alice sends the photons carrying her message to Bob~(Charlie) who can read out the message by measuring the photons in the right bases after Charlie~(Bob) publishes   his photon states; when the measurement result coincides  with that determined by Tables~\ref{Table1} and \ref{Table2}, Bob~(Charlie) receives a bit value $0$,  otherwise, he receives a bit value $1$. Subsequently, Alice publishes her information about sampling bits and Bob~(Charlie) can check the integrity of the second sequence transmission. (II) Alice measures the photons in the right bases and publishes her results over a classical channel, after she receives the basis information that Bob and Charlie use for their photon preparation; Bob~(Charlie) can then read out the message  after Charlie~(Bob) informs him of his photon states. \yyy{Although one of the communication parties performs measurement on photons carrying  sender's information, the outcome is published immediately after the measurement, which never leads to any security compromise. Therefore,  our protocol is MDI and can be immune to all side-channel attacks on practical measurement apparatuses~\cite{Xu2020Secure}.}

\begin{table}
	\centering
	\caption{Quantum state of photons remained in  Alice's node for four different cases when David announces a GHZ state  $\vert\Phi\rangle_{b_1b_20}$. $\alpha$ is the number of $\sigma_Y$ basis that Bob and Charlie use for single-photon preparation; $\beta$ is the number of states $\vert-x\rangle$ and $\vert-y\rangle$ that Bob and Charlie prepared; $\gamma_j$ is the number of state $|1\rangle$  in each $\vert\omega\rangle_j$ that are prepared in the $\sigma_Y$ basis; Case 1: $\alpha$ and $\beta$ are both even~($\alpha=0$ or $\alpha=2$, $\beta=0$~or~$\beta=2$); Case 2: $\alpha$ is even while $\beta$ is odd~({$\alpha=0$ or $\alpha=2$, $\beta=1$}); Case 3: $\alpha$ is odd while $\beta$ is even~({$\alpha=1$, $\beta=0$~or~$\beta=2$}); Case 4: $\alpha$ and $\beta$ are both odd~({$\alpha=1$, $\beta=1$}).}\label{Table1}
	\begin{tabular}{ccccc}
		\hline
		$\alpha-2\gamma_j$ & Case 1 & Case 2 & Case 3 & Case 4 \\
		\hline
		$0$ & $\vert -x \rangle$ & $\vert +x \rangle$ & $-$ & $-$  \\
		$2$ & $\vert +x \rangle$ & $\vert -x \rangle$ & $-$ & $-$  \\
		$-1$ & $-$ & $-$  & $\vert -y \rangle$ & $\vert +y \rangle$ \\
		$1$ & $-$ & $-$  & $\vert +y \rangle$ & $\vert -y \rangle$ \\
		\hline
	\end{tabular}
\end{table}

\begin{table}
	\centering
	\caption{Quantum state of photons remained in  Alice's node for four different cases when David announces a GHZ state  $\vert\Phi\rangle_{b_1b_21}$. The parameters $\alpha$, $\beta$, and  $\gamma_j$ in combination with four cases are the same as that in Table~\ref{Table1}.}\label{Table2}
	\begin{tabular}{ccccc}
		\hline
		$\alpha-2\gamma_j$ & Case 1 & Case 2 & Case 3 & Case 4 \\
		\hline
		$0$ & $\vert +x \rangle$ & $\vert -x \rangle$ & $-$ & $-$ \\
		$2$ & $\vert -x \rangle$ & $\vert +x \rangle$ & $-$ & $-$ \\
		$-1$ & $-$ & $-$ & $\vert +y \rangle$ & $\vert -y \rangle$ \\
		$1$ & $-$ & $-$ & $\vert -y \rangle$ & $\vert +y \rangle$ \\
		\hline
	\end{tabular}
\end{table}

\section {Security analysis}
So far, we have presented the three-party sender-controlled MDI-QSS protocol. An unfaithful receiver Bob, in principle, can obtain Alice's private information if he can determine the states of photons in sequence $P_A$ and hides himself in security check by introducing no disturbance. For instance, Bob can escape from {being} detected by Alice if he always announces his single-photon states after he knows the states of single photons produced by Alice, then he can obtain Alice's private information without Charlie's help  if Alice sends her photon sequence  $P_{K^{\prime\prime}}$  to Bob for message decoding and Bob measures them in proper bases.
Therefore, Bob can focus his attack on acquiring Alice's
single-photon states  using two different strategies: (I) intercept-resend attacks; (II)~teleportation-based attacks. We
will show below that both attacks can not get any useful information for Bob without being detected by the sender Alice.

(I) Intercept-resend attacks.  A direct intercept-resend attack performed by the unfaithful participant Bob can be carried out as follows:
Bob intercepts each photon of sequence $P_A$ and measures it in a randomly chosen basis $\sigma_X$ or $\sigma_Y$. He prepares a photon with the same state as her measurement outcome and sends it to David. Subsequently, Bob can completely obtain private information sent by Alice without Charlie's help if he can evade Alice's security check and Alice sends her photon sequence  $P_{K^{\prime\prime}}$   to him.
Fortunately, this intercept-resend
attack will be detected by Alice during the security check in step (3); Alice stops her message
encoding immediately after she finds the unfaithful participant and leaks none of her private message.

Alice uses single photons of sequence $P_A$ originating from $P_S$ in combination with single photon states prepared by Bob and Charlie for
security check. In principle, David can only report one of four GHZ states, when zero or two of three legitimate parties prepared their single photons in the $\sigma_Y$ basis; meanwhile, any two parties  can infer the photon state prepared by the third party. Therefore,  Bob can know Alice's photon state and then evades security check if he measures Alice's single photon in the basis she prepared it. However, Bob will disturb the outcome of David's GHZ-state analysis, if he measures Alice's photon with a conjugate basis with respect to Alice's preparation basis and sends David a photon in the corresponding state. Both cases take place with an equal probability of $1/2$ and half of the later  leads to an error outcome of David's GHZ-state analysis. Therefore, the sender Alice always asks Bob and Charlie to announce their states and determines whether even number of them prepare their photons in basis  $\sigma_Y$. Consequently, she can detect the intercept-resend attack by an average probability of $1/4$ for each single sampling process, leading to a deterministic detection of the attack with a larger sampling subset.

(II) Teleportation-based attacks. In a teleportation-based attack, Bob prepares a sequence of single photons $P_B$ and a sequence of maximally entangled photon pairs in state $\vert \varphi^-\rangle= (\vert 01\rangle -\vert 10\rangle)/\sqrt{2}$. He divides the photon pairs into two photon sequences $P_{S1}$ and $P_{S2}$ by rearranging two photons of each pair into different photon sequences without changing their orders. One photon sequence $P_{S1}$ replaces photon sequence $P_A$ and is sent to David for GHZ-state analysis, whereas the other sequence $P_{S2}$ is kept in his hand for subsequent operations. For instance, Bob can perform a collective  measurement of each photon pair ${AS2_i}$, consisting of photons ${A_i}$ and ${S2_i}$  with the same order $i$ in $P_A$ and $P_{S2}$.

When David's GHZ-state analysis of $P_{S1}$, $P_B$, and $P_C$ succeeds,
each photon in sequence $P_{S2}$ will be projected into an eigenstate of either  $\sigma_X$ or $\sigma_Y$, which is identical to that described in Tables~\ref{Table1} and \ref{Table2}. A proper two-photon measurement performed on each photon pair ${AS2_i}$ leads to two distinct results: Bob can infer the state of Alice's photon ${K_i}$ used for message encoding, if photons ${A_i}$ and ${K_i}$ come from the same entangled photon pair and are initialized to $|\varphi^-\rangle$; otherwise, Bob can only get a random result, if ${A_i}$ comes from single photon sequence $P_S$ and has no correlated photon. In the {latter} case, photons ${A_i}$ and ${S2_i}$ are both in an eigenstate of either  $\sigma_X$ or $\sigma_Y$, although the state of  ${A_i}$ is prepared by Alice and the state of  ${S2_i}$ is a collapsed state of an entangled photon pair according to Tables~\ref{Table1} and \ref{Table2}. The most efficient strategy for Bob is  measuring each ${A_i}$ with a random basis $\sigma_X$ or $\sigma_Y$ and publishing his photon state ${B_i}$ involving in GHZ-state analysis accordingly. Therefore, this teleportation-based attack will expose Bob's attack with the same probability as that of the intercept-resend attack when Alice performs security check.

\section{Multiparty sender-controlled MDI-QSS protocol}
The sender-controlled three-party MDI-QSS protocol, in principle, can be generalized to implement QSS involving ($n+1$) parties.
In a sender-controlled multiparty MDI-QSS protocol, the sender Alice can divide her message into $n$ parts and sends them to $n$ parties. The procedures of multiparty protocol is similar to that of three-party  one, except that $n\geq3$ rather than two receivers randomly prepare their photons in state  $|\pm x\rangle$ or $|\pm y\rangle$ and send them to David for ($n+1$)-photon GHZ state analysis. 
\yyy{In principle, the  ($n+1$)-photon GHZ state analysis can be achieved by quantum erasure and postselection that has been used to construct  the three-photon GHZ state analysis~\cite{Pan1998GHZanalyzer}, shown in Fig.~1. For instance, a six-photon GHZ state analysis using passive linear optics with efficiency   $1/2^5$ has been used to prepare  twelve-photon GHZ states $(|0\rangle^{\otimes6}+|1\rangle^{\otimes6})/\sqrt{2}$ out of entangled photon pairs~\cite{Zhong2018Twelve}. Furthermore, the ($n+1$)-photon GHZ state analysis can be achieved with near-unity efficiency by using nonlinear optics~\cite{Li2019Resource-efficient,qian2005universal,xia2014complete}, such as a deterministic interface between  single photons and individual electron spins~\cite{chang2014quantum,li2018gate,Song2018Photon,Qin2018Exponentially}.}
Here, we focus on the security check and briefly discuss how to complete the multiparty sender-controlled MDI-QSS protocol by encoding and decoding message.

The sender Alice prepares two photon sequences $P_A$ and $P_K$ with the same method as described in three-party protocol and sends $P_A$ to David. Meanwhile, $n$ receivers~(Bob$_1$,..., Bob$_n$) prepare $n$ photon sequences $P^{(i)}_B$ with $i=1,...,n$ and send them to David for ($n+1$)-photon GHZ state analysis. Exclusive correlations among ($n+1$) legitimate parties will be established, when David announces a successful result of his GHZ-state analyzer.

For each ($n+1$) photon with the same order in $P_A$ and $P^{(i)}_B$,  $\alpha$ $(n+1-\alpha)$ parties prepare their single photons in the $\sigma_Y$ ($\sigma_X$) basis; $\beta$~$(n+1-\beta)$ receivers  prepare their single photons in either the state $\vert-x\rangle$ or $\vert-y\rangle$~($\vert+x\rangle$ or $\vert+y\rangle$). A ($n+1$)-photon product state of these photons  can be  described  as
\begin{equation} 
	|\xi\rangle=\frac{1}{\sqrt{N}}\sum_{{j=1}}^{N}{i^{\gamma_j}(-1)^{\eta_j} \vert\omega\rangle_j},
\end{equation}
where $N=2^{(n+1)}$; $\vert\omega\rangle_j=\vert b_0\cdots b_n\rangle$ is a ($n+1$)-photon product state in the $\sigma_Z$ basis with $b_k\in\{0,1\}$ for $k=0, \cdots, n$;
$\gamma_j$ of $\alpha$ photons that are prepared  in the $\sigma_Y$ basis are in $|1\rangle$ state in each $\vert\omega\rangle_j$;
$\eta_j$ of $\beta$ photons that are prepared  in either the state $\vert-x\rangle$ or $\vert-y\rangle$ are in $|1\rangle$ state in each $\vert\omega\rangle_j$. Meanwhile, $|\xi\rangle$ can also be described as a superposition of $(n+1)$-photon states $\vert\bar{\omega}\rangle_j=\vert\bar{b}_0\cdots\bar{b}_n\rangle$ with $\bar{b}_k=1-b_k$  as follows:
\begin{equation} 
	|\xi'\rangle=\frac{1}{\sqrt{N}}\sum_{{j=1}}^{N}{i^{\alpha-\gamma_j} (-1)^{\beta-\eta_j}\vert\bar{\omega}\rangle_j}.
\end{equation}
Therefore, the state $\vert\xi\rangle$ can be rewritten  as a superposition of $\vert\omega\rangle_j$ and $\vert\bar{\omega}\rangle_j$ as follows:
\begin{equation} 
	\vert\xi''\rangle
	=\frac{1}{\sqrt{{2N}}}
	\!\sum_{{j=1}}^{N}
		(-1)^{\eta_j}i^{\gamma_j}
	[\vert\omega\rangle_j+i^{\alpha-2\gamma_j}{{(-1)^{\beta}}\vert\bar{\omega}\rangle_j})].\!\!\!\!\!
\end{equation}
The $(n+1)$-photon GHZ-state measurement performed by David can
collapse the state $\vert\xi''\rangle$ into one of $2^{(n+1)}$ GHZ states $\vert\Phi\rangle_{a_0\cdots a_n}$, i.e.,
\begin{eqnarray}  
	\vert\Phi\rangle_{a_0\cdots a_n}\!\!=\!\!\frac{1}{\sqrt{2}}[\vert0  a_0\cdots a_{n-1}\rangle+(-1)^{a_n}\vert1 \bar{a}_0\cdots \bar{a}_{n-1}\rangle],
\end{eqnarray}
where $a_k\in\{0,1\}$, $k=0, \cdots, n$, and $\bar{a}_k=1-a_k$.

In practice, David can only identify two of $2^{(n+1)}$ GHZ states, i.e., $\vert\Phi\rangle_{00\cdots 0}=(\vert00\cdots 0\rangle+\vert11\cdots 1\rangle)/\sqrt{2}$ and $\vert\Phi\rangle_{00\cdots 1}=(\vert00\cdots 0\rangle-\vert11\cdots 1\rangle)/\sqrt{2}$, when using a GHZ-state analyzer that is constituted of liner optical elements and single-photon detectors. Clearly, the success of David's GHZ-state analysis will report different GHZ states  for different $\alpha$ when no attack is involved. When $\alpha$ is even, the state $\vert\xi''\rangle$ will be specified as a superposition of $2^{n}$ $(n+1)$-photon GHZ states, in which one of each GHZ-state pair  ($\vert\Phi\rangle_{a_0\cdots a_{n-1}0}$ and $\vert\Phi\rangle_{a_0\cdots a_{n-1} 1}$)  with different phases appears. The outcome of David's GHZ-state analysis will be either $\vert\Phi\rangle_{a_0\cdots a_{n-1} 0}$ or $\vert\Phi\rangle_{a_0\cdots a_{n-1} 1}$  with a probability of $1/2^n$ for  $a_j=0$~($j=0,...,n-1$). However, when $\alpha$ is {odd}, the state $\vert\xi''\rangle$ will be specified as a superposition of $2^{(n+1)}$ $(n+1)$-photon GHZ states and $\vert\Phi\rangle_{a_0\cdots a_{n-1}  0}$ and $\vert\Phi\rangle_{a_0\cdots a_{n-1} 1}$ appear with an equal probability of $1/2^{(n+1)}$. This is the key for performing security check, since both  outside eavesdropping and participant attack will disturb the outcome of David's GHZ-state analysis and Alice can detect the disturbance after $n$  {receivers} inform her of their photon states.\yyy{Likewise, exclusive correlations are established between  photons kept in Alice's node and single photons prepared by $n$ receivers when David reports a successful outcome of his $(n+1)$-photon GHZ-state analysis using linear optics: The outcome is either $\vert\Phi\rangle_{a_0\cdots a_{n-1} 0}$ or $\vert\Phi\rangle_{a_0\cdots a_{n-1} 1}$ with an equal probability of $1/2^{(n+1)}$.}The  parties then can complete the multiparty sender-controlled MDI-QSS protocol by using similar message encoding and decoding procedures to that in three-party one, after Alice ascertains the security of photon-sequence transmission.

\section{Discussion and summary}

\yyy{In the photon-sequence transmission, we suppose that the quantum channel is ideal without any noise and thus does not change the polarization of photons transmitting over it. In practice, the channel noise will introduce depolarization and dephasing to photons that transmit directly over a quantum channel. This  changes the states of photons entering the GHZ-state analyzer, and prevents the communication parties from sharing perfect correlation shown in Tables \ref{Table1} and \ref{Table2}. Then it increases the bit--error rate of random sampling in the security check. Fortunately, the influence of  the channel noise can be suppressed by using photonic logical qubits, which are encoded in decoherence--free subspaces and are robust to channel noise~\cite{Guo2019Errorrejection,Aolita2007Quantum,Qin2015Protected,Nathan2018Open}. The GHZ state analyzer should be modified accordingly. When the bit--error rate is relatively low, forward error correction with redundant encoding can provide a  passive method to suppress the influence of quantum channel noise and has been used to perform QSDC~\cite{qi2019implementation,Massa2019twophoton,Gao2020Free-space}. For instance, a five-repetition-per-bit encoding can decrease a bit-error rate of $p_0 \leq 0.1$ to $p_1 \leq 0.0081$ for the DL04-QSDC protocol~\cite{Gao2020Free-space,Deng2004secure,Wu2019QSDC}.}


The GHZ-state analyzer situated in an ancillary node plays an important role in our sender-controlled MDI-QSS protocol and it can be partly implemented by linear optical elements and single photon detectors~\cite{Pan1998GHZanalyzer,Lu2009swap,Kop2007linear}. In practice, this type of the GHZ-state analyzer can only distinguish two GHZ states from the other $(2^n-2)$ $n$-photon GHZ states. This imposes a restriction on the transmission efficiency of the MDI-QSS protocol and decreases the efficiency exponentially with the number of receivers. Fortunately, this restriction can be lifted by using nonlinear optical elements~\cite{chang2014quantum,li2018gate,Song2018Photon}, which leads to a deterministic GHZ-state analyzer~\cite{qian2005universal,xia2014complete,Li2019Resource-efficient} and then  multiphoton postselected entanglement generation or teleportation.

For a sender-controlled MDI-QSS protocol involving $n$ receivers,
single photons with random polarization are inserted into entangled photon pairs for security check. Half of cases with desired outputs of the GHZ-state analyzer contribute to security check, in which even number of basis $\sigma_Y$ are used for single photon preparation.
A modified sender-controlled $n$-receiver MDI-QSS protocol can be obtained from the original DMDI-QSS one~\cite{gao2020deterministic},
if the sender prepares  photon pairs and single photons  with the same length and sends two photon sequences in step (2) to ancillary node for the GHZ-state analysis of $(n+2)$ photons. This protocol  equals to the original DMDI-QSS after moving one receiver of the later to sender's node and can be inherently robust to any participant attack, because it involves at least two faithful receivers, the combination of which, ascertaining security check, is out of reach of any other receivers. In principle, the modified protocol requires no basis reconciliation, each case with a desired output of the GHZ-state analyzer contributes to either security check or secret sharing. However, the efficiency of  the security check in the modified one is identical to that of the sender-controlled MDI-QSS and equals to $1/2^{(n+1)}$ when GHZ-state analyzer is implemented by linear optical elements and single photon detectors.

In summary, we have  proposed a sender-controlled MDI-QSS protocol. The sender  performs security check actively using postselected multiphoton entanglement that is generated by the GHZ-state analysis of $(n+1)$ single photons. When even number of parties prepare their single photons in basis $\sigma_Y$, an ideal outcome of GHZ-state analysis is in a subspace of $2^n$ dimensions. Attacks from either outside eavesdropper or unfaithful participant will introduce deviation from ideal outputs and can create postselected multiphoton entanglement orthogonal to the desired ones. The states of single photons prepared by the sender is the single variable that determines the ideal outputs after all receivers have published their photon states. Therefore, the sender performs as an active controller and only encodes private messages after the sender checks the security of photon-sequence transmission.
This makes our protocol useful for multiparty quantum communication networks.

\bigskip
\section*{ACKNOWLEDGMENTS}
This work was supported by  the
National Natural and Science Foundation of China (Grant No.
11904171) and the Natural Science
Foundation of Jiangsu Province (Grant No. BK20180461).


\bigskip

%

\end{document}